\documentclass[twocolumn,preprintnumbers,amsmath,amssymb,showpacs,superscriptaddress]{revtex4}

\usepackage{dcolumn}
\usepackage{bm}
\usepackage[dvipdf]{graphicx}
\DeclareGraphicsExtensions{.jpg,.pdf,.mps,.png,.eps,.ps,.EPS}
                           \begin{document}
\def\be{\begin{equation}}
\def\ee{\end{equation}}
\def\bc{\begin{center}} 
\def\ec{\end{center}}
\def\bea{\begin{eqnarray}}
\def\eea{\end{eqnarray}}
\def\bc{\begin{center}} 
\def\ec{\end{center}}
\def\bea{\begin{eqnarray}}
\def\eea{\end{eqnarray}}
\newcommand{\avg}[1]{\left\langle {#1} \right\rangle}
\def\be{\begin{equation}}
\def\ee{\end{equation}}
\title{Scale-free networks with an exponent less than two}

\author{Hamed Seyed-allaei} 
\affiliation{International School for Advanced Studies, via Beirut 4, 34014 Trieste, Italy}
\author{Ginestra Bianconi}
\affiliation{The Abdus Salam International Center for Theoretical Physics, Strada Costiera 11, 34014 Trieste, Italy}
\affiliation{INFM, UdR Trieste, via Beirut 2-4, 34014, Trieste,Italy}
\author{Matteo Marsili}
\affiliation{The Abdus Salam International Center for Theoretical Physics, Strada Costiera 11, 34014 Trieste, Italy}

\begin{abstract}
We study scale free simple graphs with an exponent of the degree
distribution $\gamma$ less than two. 
Generically one expects such extremely
skewed networks -- which occur  very frequently in systems of virtually or
logically connected units -- to have different properties than those
of scale free networks with $\gamma>2$: The number of links grows
faster than the number of nodes and 
they naturally posses 
the small world property, because the diameter increases by the logarithm of the size of the network  
and the clustering coefficient is finite. 
We discuss a simple prototype model of such networks, inspired by real
world phenomena, which exhibits these properties and allows for a
detailed analytical investigation.
 
\end{abstract}


\maketitle

There has been a recent surge of interest on the network structure
which underlie many real world phenomena \cite{dorogovtsev}. This is partly because
network's topology plays a key role in their understanding and partly
because of the ubiquity of few generic features such as the small
world property \cite{SW} and scale-free distribution of
degrees \cite{barabasi}. The latter has been observed for example in
the World Wide Web \cite{www}, Citation network \cite{cn}, Protein
Interaction Network \cite{pin}, film actors \cite{SW}, electronic
circuits \cite{ec}. Indeed in each of these systems nodes -- 
web pages or actors -- are linked -- by hyperlinks or collaboration 
in the same movie -- to a number $k$ of other nodes, which is called the 
degree of the node \footnote{By networks we mean {\em simple graphs}, those with at most one 
link between any two nodes and no tadpole. Additional features 
can be included by the generalization to weighted graphs \cite{Weights}.}, and which obeys a power 
law distribution $P(k)\sim k^{-\gamma}$.
In many cases (table \ref{my}) the exponent $\gamma$ of such a
distribution is larger than two which its occurrence has
been related to some interaction mechanism -- such as preferential
attachment \cite{barabasi} -- in simplified models.

Scale-free networks with an exponent $\gamma < 2 $ have received less
attention, despite of their widespread appearance (table \ref{my}), in
the peer-to-peer Gnutella network \footnote{Gnutella is a file sharing
  network which operates without a central server: Files are exchanged
  directly between users, using a proper software.}
\cite{gnutella_def, gnutella}, outgoing E-mails network \cite{ email},
traffic in networks \cite{traffic}, co-authorship network in high
energy physics \cite{hep} and in the network of dependency among
software packages \cite{sp,software_sole}.

The aim of this letter is to show that simple graphs with $\gamma<2$ have
markedly different properties than simple graphs with $\gamma>2$. We shall
do this first on the basis of general arguments and then using a
prototype model motivated by the above mentioned real networks. This
model reproduces all the discussed generic properties. Furthermore we
show that its generalization to a weighted network exhibits
non-trivial statistical properties.

{\em Generic properties}
 - We focus on simple graphs with uncorrelated degree distribution. 
In the ensemble of Ref. \cite{molloy_reed}, where the probability of a 
link between nodes $i$ and $j$ is $p_{i,j}=1-e^{k_ik_j/(n\avg{k})}$, 
where $\avg{k}=\sum_i k_i/n$ is the average degree, nodes with degrees
$k_i\approx \sqrt{n\avg{k}}$ cannot be considered as independent. 
The degrees of a simple scale free graph are uncorrelated only if a
structural cutoff $k_c(n)\sim \sqrt{n\avg{k}}$ is imposed in the degree 
distribution. 

Random uncorrelated networks with $\gamma<2$ differ fundamentally in their 
topology from networks with $\gamma>2$. Indeed, $\gamma<2$ implies that the average degree
increases with the system size $\avg{k}\sim n^\xi$, which means that
the total number of links grows faster than the number of nodes.
This in turn means that the cutoff $k_c(n)$ diverges
with the system size in a non-trivial manner. When $\gamma>2$ the mean 
degree $\avg{k}$ is finite and hence $k_c(n)\sim n^{1/2}$. On
the contrary, for $\gamma<2$ the divergence $\avg{k}\sim n^\xi$
implies that the structural cutoff scales with system size $n$ as
$k_c(n)\sim n^{(1+\xi)/2}$. This and the explicit calculation of
$\avg{k}$, leads to 
\begin{equation}
\xi=(2-\gamma)/\gamma.
\label{exp_identity}
\end{equation}
Correlated networks with a cutoff $k_c(n)\sim n^\chi$ which diverges faster with $n$ 
will exhibit an even faster divergence of $\avg{k}$, with $\xi=\chi(2-\gamma)$. 

Uncorrelated networks with such a broad distribution of degrees
are expected to have a high clustering coefficient. The clustering
coefficient is the ratio of number of loops of size 3
\cite{review,loops} to the number of triples of connected nodes, which
is $\sum_i k_i(k_i-1)$. So using Eq. (\ref{exp_identity}) and the fact
that $\avg{k^2}\sim k_c^{3-\gamma}$, we find a finite clustering
coefficient:
\begin{equation}\label{C}
C\sim \frac{\avg{k(k-1)}^2}{\avg{k}^3 n}\sim {\rm const}.
\end{equation}
By contrast, the same argument implies a vanishing clustering
coefficient $C\sim n^{2-\gamma}$ for $\gamma>2$.

Such a high clustering is consistent with the presence of a high
density core: Indeed a finite fraction of nodes are 
within a distance $\log\log n$ one from the other \cite{Chung-Lu}. 
Still the diameter of the network is of order $\log n$. Indeed there
is a finite number of nodes with degree $k_i=1$ and $2$ and these form
chains which connect to the core, whose length is exponentially
distributed. Hence the longest chain has length $\ell_{max}\sim \log
n$, and it dominates the behavior of the diameter. Similar arguments
were also used in reference \cite{Chung-Lu} for graphs with
$\gamma>2$.

{\em The model} - Here we study in detail a prototype model of
networks with $\gamma<2$ motivated by the real systems discussed
above. Our model is based on the idea of aggregation
\cite{amos_merge_split} and it is very similar to one recently and
independently introduced in Ref. \cite{sneppen0,sneppen1} in a
different context, and analyzed partly by Alava and Dorogovtsev
\cite{aggregation}. We show that its statistical properties can be
fully understood analytically and that they reproduce successfully the
properties observed in real world networks with $\gamma<2$.
Furthermore the model shows that, in networks with $\gamma<2$, the
statistics of strength of weighted networks can be highly non-trivial and
very different from its counterpart in networks with $\gamma>2$.
Therefore, we hope the model may serve as a starting point to
understand more complex cases as well as to address different issues,
such as the efficiency of search algorithms \cite{search}, routing,
traffic flow \cite{traffic} and transmission of infections on
peer-to-peer networks.

We consider a network of $n$ nodes and, in each time step, we perform
the following two steps:
\begin{enumerate}
\item Creation: We create a new node and connect it to a randomly chosen node. 
\item Merging: We merge two randomly selected nodes. If the two nodes
  were already connected, the corresponding link is removed. Likewise
  we remove multiple links with common neighbors of the two nodes.
\end{enumerate}

The first move is like creating a new software package, e-mail address
or running a new instance of Gnutella. The second move can be related
to merging two packages or abandoning one in the favor of another,
merging two e-mail accounts or shutting down a Gnutella client-server
and giving its load to another one. 

\begin{figure}
        \begin{center}
\includegraphics[width=6cm]{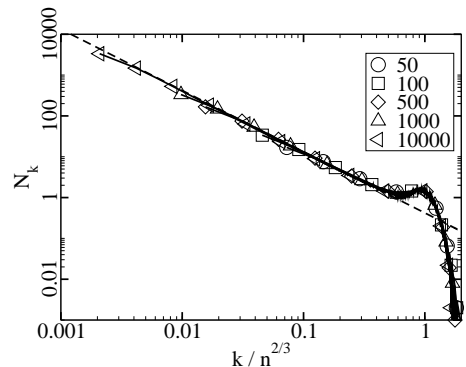}
        \end{center}
\caption{\label{scaled}
  Collapse plot of degree distribution for networks of different size.
  The dashed line corresponds to a power law with exponent $-3/2$. }

\end{figure}

The model describes a stationary network with a fixed number of nodes.
If the second process is run at a smaller rate than the first, the
model describes a growing network (see Ref. \cite{aggregation} where
a similar extension has been analyzed). Actually, to perform our
simulation, we started from a graph with a couple of nodes, then we
permitted it to grow by allowing more creation than merging until it
reached a given size. After that we merged and created nodes
sequentially to keep the number of nodes fixed and we continued it
until the system reached the stationary state of the average degree.
At that point we started taking snapshots of the network with a given
interval that was enough to give us a thousands of independent
structures. The interval between sampling was about the same time as
we had waited to reach the stationary state. We repeated the process
for different network sizes.  Results are reported in figure
\ref{scaled} and table \ref{my} compares the characteristics of our
networks to the one of real world cases.  The degree distribution
$P(k)$ follows a scaling function of the form $P(k)=
k^{-\gamma}f(k/n^\sigma)$ with $\gamma\simeq 1.5$ and $\sigma \simeq
0.67$ where $n$ indicates the total number of nodes and $k$ the degree
of the nodes. Here $f(x)$ is a scaling function with $f(x) \sim const
$ when $x \ll 1 $ and with $f(x)$ decaying faster than any power of
$x$ for $x\gg 1$.  In our model, since we have an exponent $\gamma<2$
also the total number of links $m$ follows a power-law of the form
$m=n^{\xi+1}$ with the exponent $\xi \simeq 0.33>0$, at odd with
most studied models with $\gamma>2$ for which $\xi =0$
\cite{barabasi, review}. The exponents found above 
agree perfectly with the exponent relations $\sigma=(1+\xi)/2$  and
Eq. (\ref{exp_identity}). Moreover, we found that the networks
produced by the above dynamics have the small world properties: their
diameter grows as $\log n$ with system size whereas clustering
coefficient does not decrease as $n$ increases, in agreement with
Eq. (\ref{C}).
\begin{figure}
        \begin{center}
\includegraphics[width=6cm]{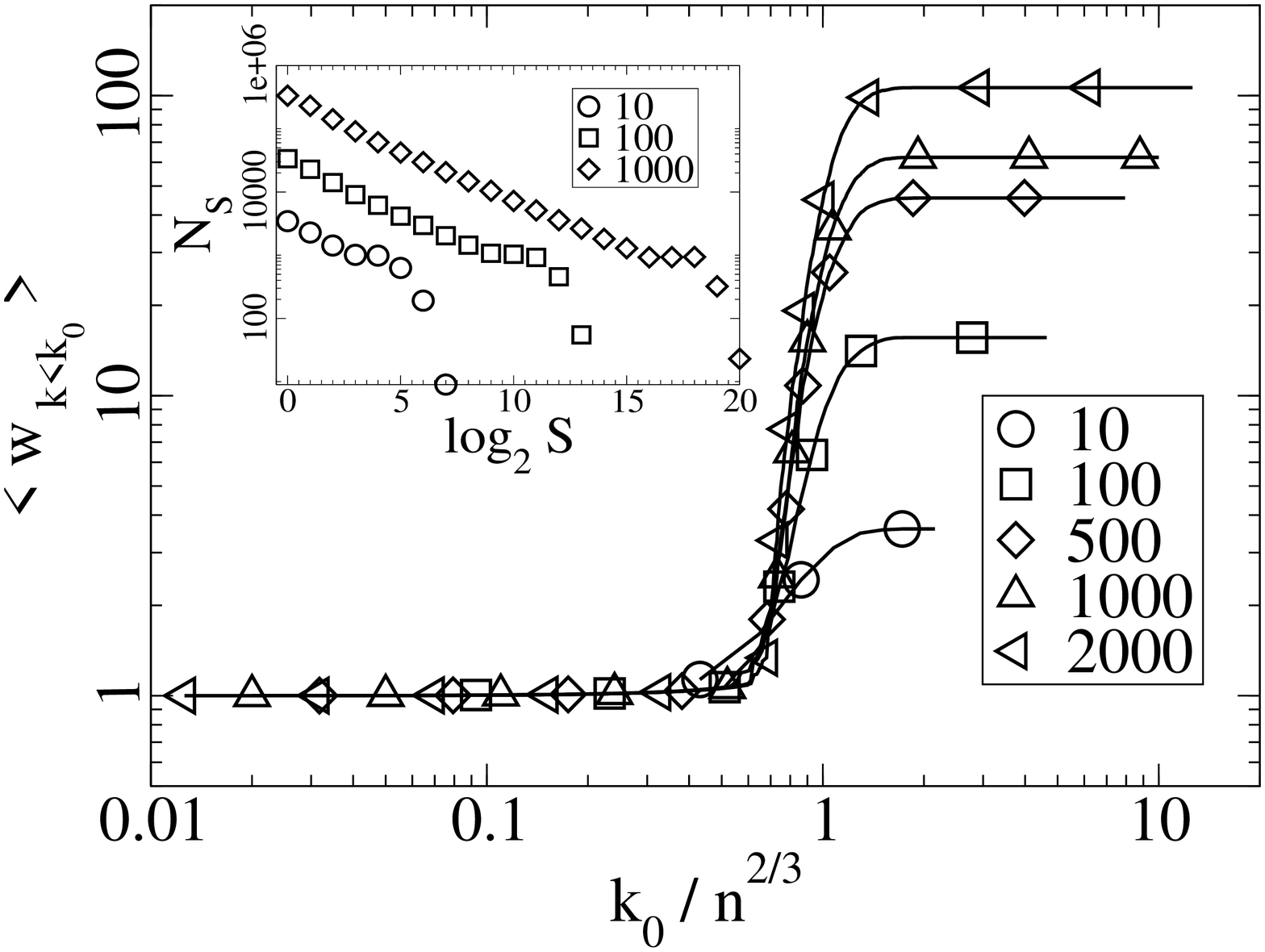}        
        \end{center}
\caption{\label{strength} 
Main figure shows the average of weight of links that are connected to nodes with a 
degree less than a given value of $k_0$ and the inset shows the histogram of strength of nodes. 
The legends show the number of nodes in each case. 
}
\end{figure}



{\em Weighted network} - It is also interesting to consider a model of
weighted networks with the above dynamics. The idea, for example, is
that if the link between two software packages $i$ and $j$ means that
package $i$ calls package $j$, it might also be interesting to keep
track of how many times $i$ calls $j$. Hence we associate a weight to
each link $ij$ and assume that it evolves according to the following
rules:
\begin{itemize}
\item A fresh link that connects a new node to the network has weight one. 
\item When merging two nodes $i$ and $j$ which are both linked to the
  same node $k$, as before we only keep one link, and its weight is
  the sum of the weights of the previous links.
\end{itemize}
In the previous example, when two software packages are merged, the
new package inherits all the calls to a third piece of software of the
merging packages. Likewise, when two e-mail account are merged, we
assume that the traffic of e-mails to a third account will be the sum
of the traffic originating from the two accounts before the merge.
This neglects the presence of complementarities, which can be an
important issue in some cases, but is the most natural way to
introduce weights in the model. Weights allow us to define the
strength of a node in the usual way \cite{Weights}, i.e. as the sum
of the weights of outgoing links.

The sum of all the weights increases when we add a node, and it
decreases when we merge two nodes that are connected; therefore, one
can expect it to reach the steady state.  This was confirmed by
simulation, which also shows that the distribution of the strengths
decays as a power-law with an exponent $1.5$. This would be consistent
with a linear relation of strength versus degree, but
Fig. \ref{strength} shows that such a relation only holds for small
$k$ and that most of the weight concentrates on high degree nodes.


{\em Analytic approach} - It is possible to shed light on these
finding and to calculate the exact value of the exponents for this
model, following similar arguments to those of Ref.
\cite{aggregation}.  We can combine the two operations above in a
single one where we replace two nodes $i$ and $j$ by two nodes of
which $i$ inherits all the links (incoming and outgoing) of both nodes
and $j$ looses all links, and acquires a new link to a randomly chosen
new node\footnote{The discussion which follows refers to an undirected
  graph, for the sake of simplicity.  The same conclusions hold for a
  directed graph if we disregard the link's direction.}. If
$k_i'$ and $k_j'$ are the degrees of the two nodes after the process,
we have
\begin{equation}\label{kprime}
  \begin{array}{c}
    {k'}_i=k_i+k_j-m_{ij}-a_{ij}\\
    {k'}_j=1 \
  \end{array}
\end{equation}
where $a_{ij}=1$ if the link $ij$ exists and $m_{ij}=\sum_{\ell}a_{i\ell}a_{j\ell}$ is the number of sites who were linked to both $i$ and $j$. Given that $i$ and $j$ are chosen at random, $m_{ij}$ and $a_{ij}$ can be regarded as random variables. The probability that the link $ij$ exists is $\avg{a_{ij}}=k_ik_j/(n\avg{k})$, likewise the average number of nodes connected to both $i$ and $j$ is 
\begin{equation}\label{mavg}
\avg{m_{ij}}=\sum_\ell\frac{k_ik_\ell}{n\avg{k}}\frac{k_\ell k_j}{n\avg{k}}=\mu k_i k_j 
\end{equation}
where $\mu={\avg{k^2}}/{n\avg{k}^2}$.
Let us now introduce the generating function for the degree distribution 
\[
\Phi(z)=\frac{1}{N}\sum_{i=1}^N E(z^{k_i}).
\]
In the stationary state, we can use Eq. (\ref{kprime}) to derive the equation
\bea
\Phi(z)&=&\frac{1}{2}{E}\left(z^{k_i+k_j-m_{ij}-a_{ij}-a_{ji}}\right)+\frac{1}{2}z\nonumber \\
&=&\frac{1}{2}\left\{z+{E}\left[z^{k_i+k_j}e^{\mu k_ik_j h(z)}\left(1+\eta k_i k_j h(z)\right)\right]\right\}
 \nonumber \label{exact}
\eea
where  $\eta=1/(n\avg{k})$, $h(z)=(1-z)/z$ 
and the last
equality hinges upon the observation that $m_{ij}$ is a Poisson
variable with mean given by Eq. (\ref{mavg}) and that $a_{ij}$ is a
random bit with $\avg{a_{ij}}=\eta k_ik_j$. Now we observe that both
$\mu$ and $\eta\to 0$ as $n\to\infty$, consequently $\Phi(z)$ can be
expanded in a power series in $\mu$ and $\eta$. The leading term
($\mu=\eta=0$) yields $2\Phi(z)=\Phi^2(z)+z$, i.e.  
\be
\Phi(z)=1-\sqrt{1-z}=\frac{1}{2\Gamma(1/2)}\sum_{k=1}^\infty \frac{\Gamma(k-1/2)}{k!} z^k; 
\ee
Therefore, for $n\to\infty$, we find 
$P(k)=\frac{1}{2\Gamma(1/2)}\frac{\Gamma(k-1/2)}{k!}\sim k^{-\gamma}$
with $\gamma=3/2$. The exponent relations derived earlier can then be
used to conclude that $\sigma=2/3$ and $\xi=1/3$. This conclusion
is also supported by a direct calculation of the next terms in the
small $\mu$ expansion. These 
finite $n$ corrections introduce a finite cutoff $k_c\sim
n^\sigma$ in the distribution, but leads to cumbersome formulas which
we will not detail here. A further way to
compute $\sigma$ comes from observing that the average of
Eq. (\ref{kprime}) in the stationary state  yields
$1=\avg{k^2}/n+\avg{k}/n$, i.e. $\avg{k^2}\sim n$. This combined with
the relation $\avg{k^q}\sim
n^{\sigma(q-\gamma+1)}$ implies $\sigma=2/3$. This shows that the
exponent relations $\sigma=(1+\xi)/2$ and Eq. (\ref{exp_identity})
-- which are valid for random graphs -- can be explicitly verified in
this model.

A simple argument also allows us to understand the statistics of
weights. Indeed at each time step, a new link with weight $w=1$ is
added. At the same time, the weight of the link between nodes $i$ and
$j$, if present, is removed. In the stationary state, then we expect
that the probability $\avg{k}/n$ of an existing link to be chosen,
times its average weight $\avg{w}$ must be equal to one. Hence
$\avg{w}\sim n/\avg{k}\sim n^{2/3}$. Unlike for the degree
distribution, we do not expect a cutoff in the distribution of
weights\footnote{This is because links are randomly drawn. Loosely
  speaking, there is no ``interaction'' term in the links' weight
  process.}.  Assuming that $P(w)\sim w^{-\eta-1}$ with $\eta<1$, we
know that $\avg{w}\sim (n\avg{k})^{1/\eta -1}$. Combining this with
$\avg{w}\sim n^{2/3}$ we find that $\eta=2/3$, in perfect agreement
with numerical simulations. Concerning the node's strength $s_i$, in
order to explain the behavior of Fig. \ref{strength} it is crucial to
observe that nodes with $k_i\ll k_c$ will have links with weights
of order one. Indeed, merge events in which nodes $i$ and $j$ share
some of their neighbors are rare if 
$\avg{m_{ij}}=(k_i/\avg{k})(k_j/\avg{k})\ll 1$, $k_i\ll k_c $ or if $k_j\ll k_c$. 
We therefore expect 
that links belonging to nodes with $k_i\ll k_c$ have weights of
order one, i.e. that $s_i\sim k_i$. For $k_i\sim k_c$ instead the
additive process of weights on links to shared neighbors becomes
relevant, eventually leading to a very broad distribution of weights
on such nodes.This is a rather non-standard situation compared to that
of most weighted networks with $\gamma>2$ \cite{Weights}. 

{\em Conclusions} - We have discussed the properties of complex scale free
networks with degree distribution exponent $\gamma<2$, which
characterizes many real systems. We have shown that these properties
are reproduced by a simple prototype model motivated by such real
systems. A key characteristic of this class of networks is that
their average degree grow with the system size, which suggests that
making a link is inexpensive.  This is indeed the case for networks of
software packages. In fact it is costly to make a package, but it is
costless to use an already existing package.  Interestingly, a
peculiarity of the model is that it involves global moves. This
requires some sort of global information exchange mechanisms, that is
not part of the network itself, that allows nodes to interact
globally. In the example of software packages, this information
exchanges happens among programmers, in fact they are responsible for
the evolution of the system and they do not exchange information only
through the system.  While both properties are likely to hold only for
open source packages, they might not apply to commercial software,
which might be expensive to link to.  A further problem is that
statistical information on commercial software dependencies is not
available.  These two features also characterize other networks: for
example, in Gnutella each node is a computer. But each link is only a
logical connection between two computers and does not require any
additional hardware. In the case of Gnutella network there are web
caches that store the information of nodes and share them with other
nodes but these caches are not considered as a part of the network
itself.  It is tempting to conjecture that the relation between these
two properties and networks with exponent $\gamma<2$ is generic. This,
applied to co-authorship network, suggests that global interaction and
information diffusion plays an essential role in establishing a dense
collaboration network.

We wish to thank S. Dorogovtsev, A. Maritan, J. Banavar 
for useful discussions. Work partially supported by EVERGROW and by EU
grant HPRN-CT-2002-00319, STIPCO.

\begin{table}
\centering
\begin{tabular}{|l|l|c|c|c|c|c|}
\hline 
\multicolumn{2}{|l|}{ Network of} & $n$ & $m$ & $\gamma$ & $C$  & $l$ \\ \hline 
\multicolumn{2}{|l|}{ Our Simulation} & 1000  & 7696  &   $3/2$    &  0.45  & 3.69 \\ \hline
\multicolumn{2}{|l|}{ Gnutella \cite{gnutella}} & 1026  & 3752 &     1.4   &  -  & 3.6  \\ 
\hline
\multicolumn{2}{|l|}{ Dependency}  &  &  &   &   &  \\ 
\multicolumn{2}{|l|}{ of software}   & 1439  & 1723  &  1.6/1.4  & 0.083   & 2.42 \\ 
\multicolumn{2}{|l|}{packages\cite{review}}   &  &  &  &  &   \\ \hline
\multicolumn{2}{|l|}{E-mails \cite{email}} &  59912 & 86300  &  1.8  &  -  & 4.95  \\ \hline
\multicolumn{2}{|l|}{Word web \cite{word_web}} &  478773 & $1.8\times10^7$  &  1.5  &  0.69  & -  \\ \hline
\multicolumn{2}{|l|}{{\small Co-authorship } } & 56627  &  9796471 &  1.2      & 0.73  & 4.0 \\ 
\multicolumn{2}{|l|}{in HEP \cite{hep}} &  &  &  & & \\ \hline \hline
\multicolumn{2}{|l|}{WWW \cite{www}} & $2\times10^8$    & $2\times10^9$  & 2.1/2.7 &  -  & 16.8  \\ \hline 
\multicolumn{2}{|l|}{Internet \cite{internet}} & 10697  & 31992  &  2.5  & 0.035 & 3.31  \\ \hline
\multicolumn{2}{|l|}{PIN \cite{pin}} & 2115  &  2240 &  2.4 & 0.072 & 6.8 \\ \hline
\multicolumn{2}{|l|}{Citation \cite{cn}} & 783339  & 6716198  &   3.0   &  - & -  \\ \hline
\multicolumn{2}{|l|}{Actors \cite{SW} }& 449913  & 25516482  &  2.3  & 0.20 & 3.48   \\ \hline
\multicolumn{2}{|l|}{Electronic }  & 24097  & 53248  &  3.0      & 0.010 & 11.05 \\ 
\multicolumn{2}{|l|}{Circuits \cite{ec}}  &   &   &     &  &  \\ \hline
\end{tabular} 

\caption{\label{my}  Results of our simulation and its comparison to some empirical observations.  
In the case of directed network the exponents is shown in the form of in/out. Here the total number of nodes, links, the exponent, clustering coefficient, mean of shortest paths  are represented by  $n$, $m$, $\gamma$, $C$, $l$.}
\end{table}

\bibliography{hamed_graph}

\end{document}